\documentclass[aps,pra,notitlepage,twocolumn]{revtex4-1}
\usepackage{amsmath,amssymb,graphicx}

\begin{document}

\newcommand{\bra}[1]{\langle #1|}
\newcommand{\ket}[1]{| #1\rangle}
\newcommand{\tr}[1]{{\rm Tr}\left[ #1 \right]}
\newcommand{\av}[1]{\langle #1 \rangle}
\renewcommand{\k}{{\bf k}}
\newcommand{\x}{{\bf r}}
\newcommand{\ketbra}[2]{|#1\rangle\!\langle#2|}

\title{Detection of Bell correlations at finite temperature\\ from matter-wave interference fringes}

\author{A. Niezgoda$^1$, J. Chwede\'nczuk$^1$, L. Pezz\'e$^2$ and A. Smerzi$^2$}
\affiliation{$^1$Faculty of Physics, University of Warsaw, ul. Pasteura 5, PL--02--093 Warszawa, Poland\\
  $^2$QSTAR, INO-CNR and LENS,  Largo Enrico Fermi 2, 50125 Firenze, Italy}

\begin{abstract}
  We show that matter-wave interference fringes formed by two overlapping atomic clouds can yield information about the 
  non-local Bell correlations. 
  To this end, we consider a simple atomic interferometer, where the clouds are released from the double-well potential and the relative phase is estimated from the density fit to this interference pattern.
  The Bell correlations can be deduced from the sensitivity of the phase obtained in this way.
  We examine the relation between these two quantities for a wide range of ground states of the double-well, scanning through the attractive and
  the repulsive interactions. The presented analysis includes the effects of finite temperature, when excited states are thermally occupied. 
  We also consider the impact of the spatial resolution of the single-atom detectors and the fluctuations of the energy mismatch between the wells. 
  These results 
  establish a link between the fundamental (non-locality) and application-oriented (quantum metrology) aspects of entanglement.
\end{abstract}

\maketitle

{\it Introduction.} The interference pattern obtained after time-of-flight imaging~\cite{AndrewsSCIENCE1997} 
is a direct probe of phase coherence in ultracold gases~\cite{BlochRMP2008, InguscioBOOK}.
Interference effects have been used to probe the superfluid to Mott-insulator quantum phase transition in an optical lattice~\cite{GreinerNATURE2002a}, 
detect the Berezinskii-Kosterlitz-Thouless transition in two-dimensional quantum gases~\cite{HadzibabicNATURE2006}
demonstrate collapse and revivals of phase coherence~\cite{GreinerNATURE2002b},
reveal the quantum statistics of bosonic~\cite{FollingNATURE2005} and fermionic~\cite{RomNATURE2006} atoms via Hanbury-Brown and Twiss correlations, 
measure the effective spin length of a condensate in a double-well systems~\cite{EsteveNATURE2008,BerradaNATCOMM2013}, and 
study the evolution of the relative phase in Josephson experiments~\cite{CataliottiSCIENCE2001, AlbiezPRL005, SpagnolliPRL2017}.
Moreover, interference of condensates released from an optical lattice has been used to demonstrate mode entanglement~\cite{CramerNATCOMM2013} 
and is a tool to extract the Renyi entropy in optical lattices~\cite{IslamNATURE2015}.
Multipartite entanglement between atoms in a double-well potential~\cite{pezze2009entanglement,hyllus2012fisher,toth2012multipartite,PezzePRA2005} can be detected by the phase uncertainty otained
in repeated interference experiments with Bose-Einstein condensates~\cite{ChwedenczukNJP2012}.    

In this manuscript, we show that interference fringes formed by two weakly-linked Bose-Einstein condensates---initially trapped in a double-well potential and 
overlapping after free expansion---can reveal Bell correlations between atoms through the sole analysis of the one-body density distribution.
We relate the quantum phase sensitivity obtained from the analysis of the density pattern after the matter-wave expansion 
to the criteria for witnessing Bell correlations based on the first- and second-order 
correlation function of bosons derived in Ref.~\cite{TuraSCIENCE2014} (see also Refs.~\cite{PelissonPRA2016, TuraPRX2017, FadelQ2018, BaccariARXIV,WasakPRL2018} for recent studies).
Bell's correlations---recently observed in atomic ensembles~\cite{SchmiedSCIENCE2016,WagnerPRL2017,EngelsenPRL2017,shin2018strong}---are a strong form of entanglement necessary 
to violate Bell's inequalities~\cite{HorodeckiRMP2009,BrunnerRMP2014}
and eventually demonstrate non-locality of quantum mechanics~\cite{EPR1935, BellPHYS1964,bell_rmp}. 
Differently from existing experimental studies~\cite{SchmiedSCIENCE2016,WagnerPRL2017,EngelsenPRL2017}, where
Bell correlations between neutral atoms have been observed using internal degrees of freedom (atomic hyperfine states)
and the measurement of the composite spin vector required a number of manipulations of the system,
in our case Bell correlations are witnessed using external degrees of freedom and by the observation of the interference pattern.
We demonstrate that Bell correlations in the Bose-gas is naturally present in the ground state of the system without requiring, in the case of attractive interaction, 
further manipulation.  
Our study takes into account relevant experimental imperfections such as finite temperature, 
finite spatial resolution of single-atom detectors, and the fluctuations of the energy mismatch between the wells of the double well potential.

{\it Model and methods.}  We consider an ultracold Bose gas trapped in a 
double-well potential and with tunable interparticle interaction~\cite{SpagnolliPRL2017, TrenkwalderNATPHYS2016, PezzeRMP2018}. 
For a sufficiently high tunneling barrier and relatively weak interaction, the system can be described 
in a two-mode approximation (see Ref.~\cite{PezzeRMP2018} for a review). 
In this case, the bosonic field operator is
$\hat{\Psi}(\x,t) = \psi_{a}(\x,t)\hat a+\psi_{b}(\x,t)\hat b$, 
where $\hat a$ and $\hat b$ annihilates a particle in the left or right well of the potential, respectively, 
$\psi_{a,b}(\x,t)$ are the corresponding (real) spatial wavefunctions centered around the minima of the double-well trap
satisfying $\int\! d\x\, \psi_{a,b}^2(\x,t) =1$ and $\int\! d\x\,\psi_{a}(\x,t) \psi_{b}(\x,t) = 0$.
The normalization condition sets $\int\! d\x\av{\hat\Psi^\dagger(\x,t)\hat\Psi(\x,t)} = \av{\hat{a}^\dag\hat a + \hat{b}^\dag\hat b} =  N$, 
where $N$ is the conserved total number of particles. 
Within this two-mode approximation, the system can be described by the 
bosonic Josephson junction Hamiltonian
\begin{align}\label{eq.ham}
  \hat H = -\hat J_x+ \frac{\Lambda}{N} \hat J^2_z + \delta \hat J_z,
\end{align}
where 
\begin{align}  \label{eqs.angular}
  \hat{J}_x = \frac{\hat{a}^\dagger\hat{b}+ \hat{b}^\dagger\hat{a}}{2}, \,\,\,
  \hat{J}_y = \frac{\hat{a}^\dagger\hat{b}-\hat{b}^\dagger\hat{a}}{2i}, \,\,\,
  \hat{J}_z = \frac{\hat{a}^\dagger\hat{a}-\hat{b}^\dagger\hat{b}}{2}
\end{align}
is the triad of the angular momentum operators.
In Eq.~(\ref{eq.ham}) the parameters are re-scaled to the Josephson tunneling energy $E_J$.
And so, $\Lambda=U/E_J$ rules the competition between interaction and tunneling---it can be positive or negative, depending wether the interaction strength is repulsive or attractive, respectively. 
The parameter $\delta$ depends on the energy mismatch between the two wells.
It should be noticed that, for $\delta=0$, the ground state of Eq. (\ref{eq.ham}) undergoes a second-order quantum phase transition at $\Lambda=-1$
between a paramagnetic (at $\Lambda>-1$) and a ferromagnetic ($\Lambda<-1$) phase~\cite{PezzeRMP2018,UlyanovPR1992,VidalPRA2004,ZinEPL2008,BuonsantePRA2012}.
For a fixed value $\Lambda<-1$ and tuning the energy mismatch $\delta$, the system has a first order quantum phase transition 
with a discontinuous jump of the population imbalance. 
First- and second-order quantum phase transitions in this system have been experimentally observed in Ref.~\cite{TrenkwalderNATPHYS2016}. 

After the preparation of the condensate in the double-well (we consider the zero- as well as finite-temperature cases), 
we give a phase shift $\varphi$ between the two modes.
This is obtained by applying an energy imbalance $\delta_\varphi$ between the two modes for a time $t_\varphi$ such that
$\varphi = \delta_\varphi t_\varphi$ (we set $\hbar\equiv1$), assuming that the effect of tunneling and interaction is negligible during the 
phase acquisition time, $E_J t_\varphi, U t_\varphi \ll \varphi$..
We also assume that $\varphi$ is reproducible in repeated independent experiments.  
After phase acquisition, the trap is switched off and the wavefunctions, initially localized in the two wells, expand and overlap. 
In the far field (namely, for a sufficiently long time of flight such that the wavefunctions fully overlap), the (normalized) spatial density 
is given by 
\begin{align}\label{eq.dens}
  \varrho(\x, t_{\rm f};\varphi)=\frac{\av{\hat\Psi^\dagger(\x,t_{\rm f})\hat\Psi(\x,t_{\rm f})}}N=1+\nu\cos(\k\cdot\x+\varphi).
\end{align}
Here, $\k=2\x_0/\frac{\hbar t_{\rm f}}m$, $2\x_0$ is the vector pointing from the center of one well to the other, 
$t_{\rm f}$ is the time of flight, $m$ is the mass of a single atom, and
\begin{align}
  \nu=\frac2N|\av{\hat J_x}|
\end{align}
is the visibility of the interference fringes (notice that $\nu \leqslant 1$).
We take $\av{\hat J_y}=0$ throughout the text---this is satisfied for all the ground states of the double-well potential, 
as well as for all eigenstates of the Hamiltonian (\ref{eq.ham}). This condition is altered neither by the thermal fluctuations nor other sources of noise considered below.

The phase $\varphi$ can be estimated by measuring the 
position of atoms, and then fitting the one-body density 
$\varrho(\x, t_{\rm f};\varphi_{\rm est})$, Eq.~(\ref{eq.dens}), using the least-squares method, with $\varphi_{\rm est}$ as a free parameter. 
This gives an estimator $\varphi_{\rm est}$ of $\varphi$ which is unbiased~\cite{ChwedenczukNJP2012}, $\overline{\varphi_{\rm est}} = \varphi$ 
as the number of experiments tends to infinity, where the over-line denotes statistical averaging.
The variance of this estimator, taken as measure of sensitivity of the double-well interferometer, is equal to~\cite{ChwedenczukNJP2012}
\begin{align}\label{eq.fit}
  \Delta^2\varphi_{\rm est}=\frac1{N}\left[\xi_\phi^2+\frac{\sqrt{1-\nu^2}}{\nu^2}\right],
\end{align} 
where $\xi_\phi^2=N \av{\hat J_y^2} / \langle\hat J_x\rangle^2$ is the spin-squeezing parameter \cite{PezzeRMP2018}.
The condition $\xi_\phi^2<1$ implies squeezing of the $\hat{J}_y$ spin component, that is generally indicated as phase-squeezing.
We introduce the parameter 
\begin{align}\label{eq.A}
  \mathcal{A}\equiv N\Delta^2\varphi_{\rm est}-1 =\xi_\phi^2+\frac{\sqrt{1-\nu^2}-\nu^2}{\nu^2}.
\end{align}
Sub shot-noise sensitivity  i.e., $\Delta^2\varphi_{\rm est}<\frac1{N}$ in the estimation of $\varphi$ is equivalent to $\mathcal A<0$.

In this manuscript we relate Eq.~(\ref{eq.fit}) to the witness of 
Bell's correlations introduced in Refs.~\cite{TuraSCIENCE2014, SchmiedSCIENCE2016}, namely
$B(\theta)\equiv \langle \hat{ B}(\theta) \rangle$, where 
the operator
\begin{align} \label{eq.bell}
  \hat{ B}(\theta)&=2N\cos^2\frac\theta2-4\hat J_1+\nonumber\\
  & + 8\sin^2\frac\theta2\left[-\hat J_1\sin\left(\frac\theta2\right)+\hat J_2\cos\left(\frac\theta2\right)\right]^2
\end{align}
contains only one- and two-body operators.
Here the subscripts ``1'' and ``2'' denote a pair of orthogonal directions. 
The system contains Bell correlations if $B(\theta)<0$.

To relate $\Delta^2\varphi_{\rm est}$---or equivalently the parameter $\mathcal A$---to the witness of Bell correlations,
it is convenient to apply the transformation $\hat{J}_1 = \hat{J}_x \cos\left(\tfrac{\theta}{2}\right) - \hat{J}_y \sin\left(\tfrac{\theta}{2}\right)$ and 
$\hat{J}_2 = \hat{J}_x \sin\left(\tfrac{\theta}{2}\right) + \hat{J}_y \cos\left(\tfrac{\theta}{2}\right)$
to Eq.~(\ref{eq.bell}), giving~\cite{SchmiedSCIENCE2016} 
\begin{align}\label{eq.belltheta}
  B(\theta)&=2N\cos^2\frac\theta2-4 \langle \hat J_x \rangle \cos\left(\frac\theta2\right)+8\sin^2\left(\frac\theta2\right) \langle \hat J_y^2 \rangle.
\end{align}
This equation can be minimized analytically with respect to $\theta$ and
the minimum is reached for
\begin{align}\label{eq.ineq}
  \cos\left(\frac{\theta_0}2\right)  =\frac{\nu}{2( 1-\xi_\phi^2\nu^2 ) }.
\end{align}
Replacing Eq.~(\ref{eq.ineq}) into Eq.~(\ref{eq.belltheta}) we obtain that Bell's correlation are witnessed when  
\begin{align}\label{eq.bell}
  \mathcal{B}  \equiv B(\theta_0) = \xi^2_{\phi}+\frac{\sqrt{1-\nu^2}-1}{2\nu^2} < 0.
\end{align}
Finally, the parameter $\mathcal A$ from Eq.~(\ref{eq.A}) and $\mathcal B$ from (\ref{eq.bell}) are related by a simple function of the fringe visibility:
\begin{align}\label{eq.rel}
  \mathcal B=\mathcal A+f(\nu), 
\end{align}
where $f(\nu)=1-\frac{\sqrt{1-\nu^2}+1}{2\nu^2}$ and $\mathcal{A}$ is given by Eq.~(\ref{eq.A}).
Thus the fit of the one-body function to the interference pattern brings not only the knowledge about the phase and its sensitivity~\cite{ChwedenczukNJP2012}, but also
about the Bell correlations in the system. 

\begin{figure}[t!]
  {\includegraphics[width=\columnwidth]{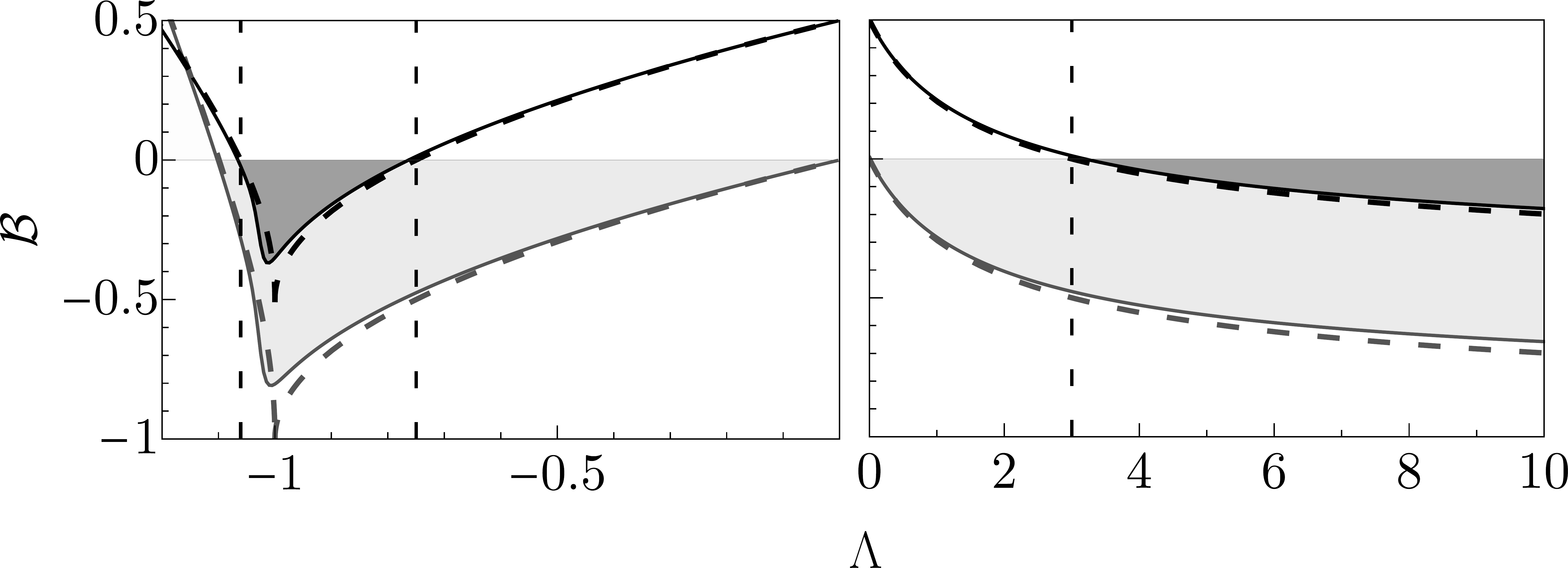}}
  \caption{The numerical (solid lines) and analytical (dashed lines) results for coefficients $\mathcal A$ (gray lines) and $\mathcal B$ (black lines) as a function of the interaction strength 
    $\Lambda$ for both attractive (left) and repulsive (right) interactions,
    $T = 0$, $\delta = 0$ and $N=1000$. 
    The shaded areas indicate the regimes $\mathcal A \leqslant 0$ (lighter area) and $\mathcal B \leqslant 0$ (darker area). Vertical lines for $\Lambda = \frac{-3}{2\sqrt{2}}, -\frac{3}{4}, 3$ 
    indicate the analytical solution for $\mathcal{B}=0$. }
  \label{Fig1}
\end{figure}

{\it Noiseless case.} Let us consider first the noiseless case where the system is prepared in the ground state of Eq.~(\ref{eq.ham}) with $\delta=0$.
For attractive interaction, the ground state is phase-squeezed for $-(1+\sqrt{5})/2<\Lambda<0$~\cite{GabbrielliSR2018}. 
In a semiclassical approximation~\cite{ShchesnovichPRA2008,JuliaDiazPRA2012}, valid for $N \gg 1$,  
we have $\xi_\phi^2 \approx \sqrt{1+\Lambda}$ and $\nu \approx 1$ for $-1 \lesssim \Lambda \leqslant 0$, so that
\begin{align} \label{AB1}
  \mathcal A \approx \sqrt{1+\Lambda}-1, \qquad  \mathcal B \approx \sqrt{1+\Lambda}-1/2.
\end{align}
The  condition $\mathcal B <0$ reduces to $\xi_\phi^2 < 1/2$, which is achieved for $\Lambda < -3/4$.
For $-(1+\sqrt{5})/2<\Lambda<-1$, we have $\xi_\phi^2 \approx  |\Lambda| \sqrt{\Lambda^2-1}$ and $\nu \approx 1/|\Lambda|$, giving
\begin{align} \label{AB2}
  \mathcal A \approx 2|\Lambda| \sqrt{\Lambda^2-1}-1, \quad  \mathcal B \approx \frac{3}{2}|\Lambda| \sqrt{\Lambda^2-1}-\frac{\Lambda^2}{2}.
\end{align}
The  condition $\mathcal B <0$ gives $\Lambda > -3/(2\sqrt{2})$.
The analytical expressions (\ref{AB1}) and (\ref{AB2}) are quite accurate for sufficiently large $N$, except around $\Lambda = -1$
where the approximations used to derive these expressions break down~\cite{GabbrielliSR2018}.
For repulsive interaction, $\Lambda>0$, the ground state of Eq.~(\ref{eq.ham}) is number-squeezed, 
namely $\xi_N^2 = N (\Delta \hat{J}_z)^2/\langle \hat{J}_x \rangle^2 < 1$ \cite{PezzeRMP2018,EsteveNATURE2008,BerradaNATCOMM2013}.
A further rotation of an angle $\pi/2$ around the $x$ axis is required in order to 
transform number-squeezing onto phase-squeezing, giving $\nu \approx 1$ and 
\begin{align} \label{AB3}
  \mathcal A \approx \frac{1}{\sqrt{1+\Lambda}}-1, \quad \mathcal B \approx \frac{1}{\sqrt{1+\Lambda}}-\frac{1}{2},
\end{align}
for $0 \leqslant \Lambda \ll N^2$.
This rotation is achieved, for instance, by a quench of the tunneling for a time $t_{\pi/2} E_J = \pi/2$ such that
$t_{\pi/2} U \ll 1$ is negligible. 
The condition to observe Bell correlations, $\mathcal B<0$, is $\xi^2_\phi<1/2$, that is fulfilled for $\Lambda > 3$.
In Fig.~\ref{Fig1} we plot $\mathcal A$ (solid line) and $\mathcal B$ (dashed line) as a function of $\Lambda$ and for $N=1000$:
analytical results are well reproduced by the numerical calculation (obtained via exact diagonalization). 
The slight discrepancy between the numerical results and the analytical prediction for $\Lambda>0$ is due to the finite 
atom number and the assumption $\nu \approx 1$ obtained to derive Eq.~(\ref{AB3}).

\begin{figure}[t!]
  {\includegraphics[width=\columnwidth]{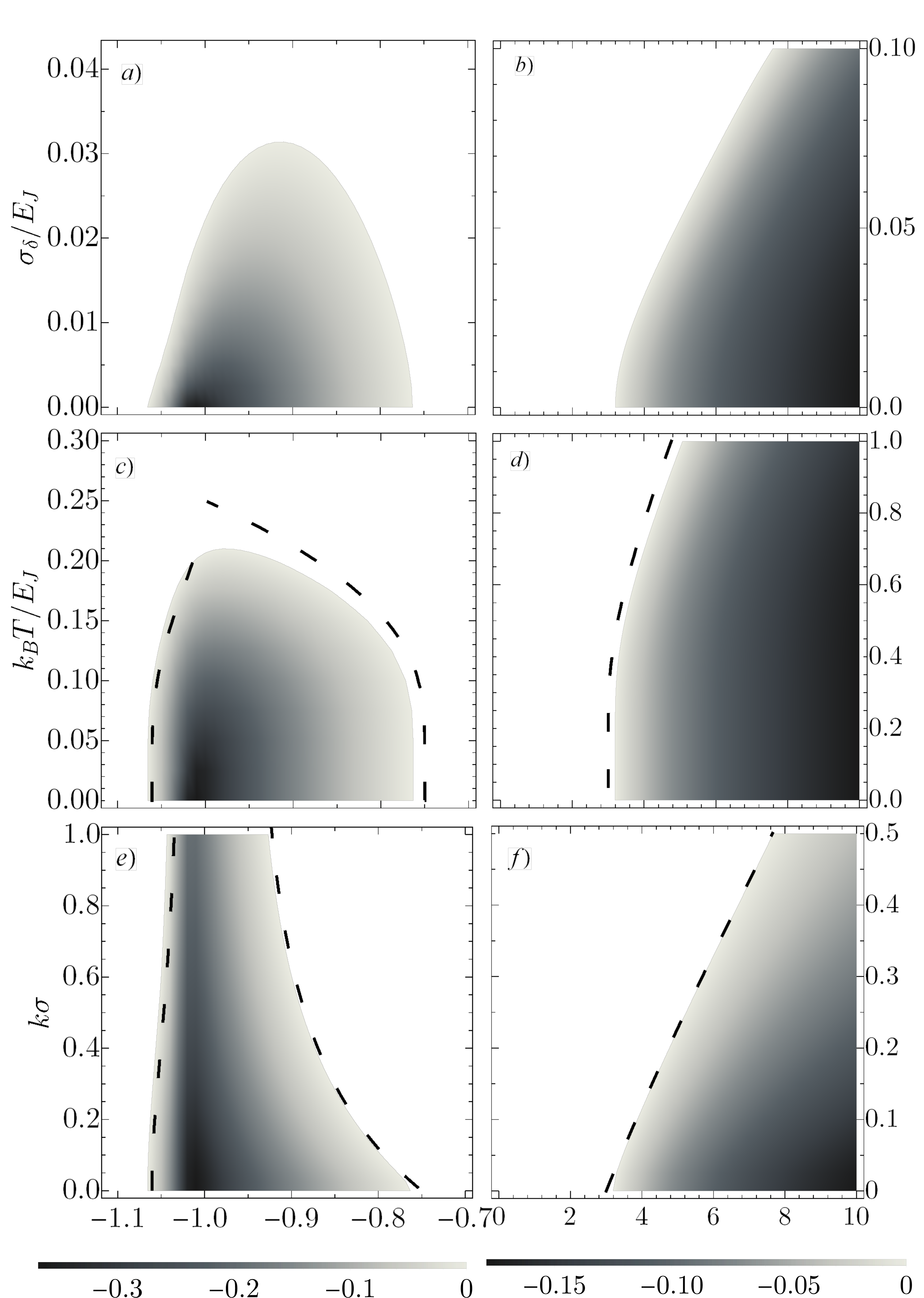}}
  \caption{Parameter regions where the condition $\mathcal{B} < 0$ (shaded area) is fulfilled. 
    Panels (a) and (b) show the effect of energy imbalance (modeled as a normal distribution with width $\sigma_\delta$) at $T=0$;
    (c) and (d) show the finite temperatures $T$ case, for $\sigma_\delta = 0$;
    (e) and (f) show the effect of finite resolution $\sigma$ of spatial detection of the atoms, at $T=0$ and $\sigma_\delta=0$.
    The dashed lines in panels $c - f$ are analytic predictions for $\mathcal{B}=0$ 
    obtained from Eq.~(\ref{eq.btsigma}). In all panels $N=1000$.
  }
  \label{Fig2}
\end{figure}

{\it Noisy case.} In a more realistic scenario, we include the noise coming from three different sources. 
First, we take into account non-vanishing energy imbalance between the wells, i.e., $\delta\neq0$, 
which is one of the leading sources of noise in current double well experiments~\cite{SpagnolliPRL2017, TrenkwalderNATPHYS2016}. 
We model shot-to-shot fluctuations of $\delta$ with a Gaussian distribution of width $\sigma_\delta$, such that the quantum state of the system if given by  
a density matrix
\begin{align}\label{eq.imb}
  \hat\varrho_{\sigma_\delta, \Lambda}=\mathcal N\int\!\! d\delta\, e^{-\frac{\delta^2}{2 \sigma_\delta^2}}\ketbra{\Psi_{\delta,\Lambda}}{\Psi_{\delta, \Lambda}},
\end{align}
where $\mathcal N$ is the normalization constant and $\vert \Psi_{\delta, \Lambda} \rangle$ is the ground state of 
Eq.~(\ref{eq.ham}) for fixed values of $\Lambda$ and $\delta$. 
We calculate $\mathcal A$ and $\mathcal B$ for Eq.~(\ref{eq.imb}) and plot the result in Fig.~\ref{Fig2}(a) and (b). 
There, we show the regions of $\mathcal B<0$:
the darker the shade of gray, the more negative value of $\mathcal B$. 
With growing $\sigma_\delta$ the range of values of $\Lambda$
for which the Bell correlations are witnessed by $\mathcal B$ shrinks, and the effect is much more pronounced for attractive interactions. The regions of SSN sensitivity, $\mathcal A<0$
shrinks proportionally, according to Eq.~(\ref{eq.rel}). 

Next, we consider the effects of non-zero temperature. 
First, we construct the density matrix at thermal equilibrium for the Hamiltonian (\ref{eq.ham}), namely
\begin{align}\label{eq.mat}
  \hat\varrho_{\rm th}=\frac1{\mathcal Z}\sum_{n=0}^N \ket{\Psi_n} \bra{\Psi_n} e^{-\beta E_n},
\end{align}
where $\mathcal Z$ is the partition function,
$\hat H\ket{\Psi_n}=E_n\ket{\Psi_n}$ and $\beta=E_J/(k_B T)$ (where $k_B$ is the Boltzmann constant).
The coefficients $\mathcal{A}$ and $\mathcal{B}$ are obtained from a calculation of the relevant spin moments  
using $\hat\varrho_{\rm th}$, for instance
$\av{\hat{J}_y^2}_{\rm th} = {\rm Tr} \big[ \hat\varrho_{\rm th}\hat J_y^2 \big] = \sum_{n=0}^N \frac{e^{-\beta E_n}}{\mathcal Z}\bra{\Psi_n}\hat J_y^2\ket{\Psi_n}$.
Numerical results are shown in Figs.~\ref{Fig2}(c) and (d).
The dashed lines in these panels give $\mathcal{B}=0$ are obtained from an analytical 
expression for the spin-squeezing parameter~\cite{GabbrielliSR2018}, valid for sufficiently large $N$,
\begin{align}\label{eq.xi}
  \xi_\phi^2(T)=\left\{
  \begin{array}{ll}
    \vert\Lambda\vert\sqrt{\Lambda^2-1}\coth\left({\frac{\beta\sqrt{\Lambda^2-1}}{2}}\right),&\Lambda < -1,\\
    \sqrt{1 + \Lambda}\coth\left({\frac{\beta\sqrt{1 + \Lambda}}{2}}\right),& -1<\Lambda < 0,\\
    \frac{1}{\sqrt{1 + \Lambda}}\coth\left({\frac{\beta\sqrt{1 + \Lambda}}{2}}\right),&\Lambda>0,
  \end{array}
  \right.
\end{align}
and assuming $\nu \approx 1$ and reproduce quite accurately the numerical results. 

Finally, the third effect that we consider is that of the finite resolution in the detection of the atoms. 
To model this effect we convolute the density Eq.~(\ref{eq.dens}) with a gaussian probability of detecting an atom at position $\x$ given its true position $\x'$, namely
\begin{align}
  \tilde\varrho(\x, t_f;\varphi)&=\frac1{(\sqrt{2\pi}\sigma)^3}\int d\x ~ e^{-\frac{(\x-\x')^2}{2\sigma^2}}\varrho(\x,t_f;\varphi)=\nonumber\\
  &=1+\tilde\nu\cos(\k\cdot\x+\varphi),
\end{align}
where $\tilde\nu=\nu e^{-\frac12k^2\sigma^2}$ is a blurred visibility. 
Using $\tilde\varrho$, we calculate
the sensitivity $\Delta^2\varphi_{\rm est}$ and, from Eq.~(\ref{eq.rel}), we obtain the expression for the Bell witness, i.e., 
\begin{align}\label{eq.btsigma}
  \mathcal B(T,\sigma)=\xi_\phi^2(T)+\frac{\sqrt{1 - \tilde\nu^2 }-1}{2\tilde\nu^2}.
\end{align}
Fig.~\ref{Fig2} (d)-(e), we show the region of parameters for which $\mathcal B(0,\sigma) \leqslant 0$, while the dashed line is the analytical prediction for $\mathcal{B}=0$.

{\it Conclusions.} In this paper we have shown that the observation of the one-body density distribution of atoms released from 
a double-well potential and forming an interference pattern can witness the existence of non-local Bell correlations is this system. 
This is achieved by a precise link between the precision of phase estimation obtained from a fit of the density to this pattern with 
the Bell coefficient $\mathcal B$ introduced in Refs.~\cite{TuraSCIENCE2014,SchmiedSCIENCE2016}. 
We have analyzed the relation between these two quantities for the bosonic Josephson junction Hamiltonian, with
both for attractive and repulsive interaction, including the effects of finite temperature, energy imbalance between the two wells and 
finite detection efficiency.
Our results provide an experimentally feasible method of detecting the Bell correlations and establish
a link between the fundamental and the application-oriented aspects of entanglement.

{\it Ackonwledgements.}
We thank M. Fattori for comments and discussions.
A. N. and J.Ch. acknowledge the support of Project no.
2017/25/Z/ST2/03039 funded by the National Science Centre, Poland under
QuantERA programme. This work is also supported by the QuantERA ERA-NET
Cofund in Quantum Technologies  projects ``TAIOL'' and ``CEBBEC''.

\end{document}